\documentclass[aps,prd,preprintnumbers,nofootinbib]{revtex4}
\usepackage{epsfig}

\def\tr{\textrm{tr}}
\def\Tr{\textrm{Tr}}
\def\Str{\textrm{Str}}
\def\det{\textrm{det}}

\def\etal{{\it et al.}}

\def\spose#1{\hbox to 0pt{#1\hss}}
\def\ltapprox{\mathrel{\spose{\lower 3pt\hbox{$\mathchar"218$}}
 \raise 2.0pt\hbox{$\mathchar"13C$}}}
\def\gtapprox{\mathrel{\spose{\lower 3pt\hbox{$\mathchar"218$}}
 \raise 2.0pt\hbox{$\mathchar"13E$}}}
\def\inapprox{\mathrel{\spose{\lower 3pt\hbox{$\mathchar"218$}}
 \raise 2.0pt\hbox{$\mathchar"232$}}}

\preprint{UW/PT 03-23}
\begin{document}

\title{Unphysical Operators in Partially Quenched QCD}

\author{Stephen R. Sharpe}
\email{sharpe@phys.washington.edu}
\affiliation{Physics Department, University of Washington,
Seattle, WA 98195-1560}
\author{Ruth S. Van de Water}
\email{ruthv@u.washington.edu}
\affiliation{Physics Department, University of Washington,
Seattle, WA 98195-1560}

\date{\today}

\begin{abstract}
We point out that the chiral Lagrangian describing pseudo-Goldstone
bosons in partially quenched QCD has one more
four-derivative operator than that for unquenched QCD with three flavors.
The new operator can be chosen to vanish in
the unquenched sector of the partially quenched theory.
Its contributions begin
at next-to-leading order in the chiral expansion.
At this order it contributes only to unphysical scattering
processes, and we work out some examples.
Its contributions to pseudo-Goldstone properties 
begin at next-to-next-to-leading order, and we determine 
their form.
We also determine all the zero and two derivative operators
in the $O(p^6)$ partially quenched chiral Lagrangian, finding
three more than in unquenched QCD, and use these to give
the general form of the analytic next-to-next-to-leading
order contributions to the pseudo-Goldstone mass and decay constant.
We discuss the general implications of such additional operators
for the utility of partially quenched simulations.
\end{abstract}

\maketitle

\section{Introduction}
\label{sec:intro} 
Chiral perturbation theory ($\chi$PT) allows analytic calculations of
low-energy QCD processes, results of which are given in terms of a
number of undetermined constants, e.g. the
Gasser-Leutwyler coefficients.  These low-energy constants are
fundamental QCD parameters that govern physical properties such as
masses and scattering amplitudes.
Lattice QCD provides a method for determining them from
first principles, 
as long as simulations are done at small enough quark masses that
$\chi$PT (typically at next-to-leading order) is a good approximation.
In practice, however, simulating light dynamical quarks is
computationally expensive, and there is a limited range of
quark masses where both lattice simulations are feasible and $\chi$PT
is applicable.

\begin{figure} \epsfysize=2.1in
\vspace{.2in}\hspace{-.0in}\rotatebox{0}{\leavevmode\epsffile{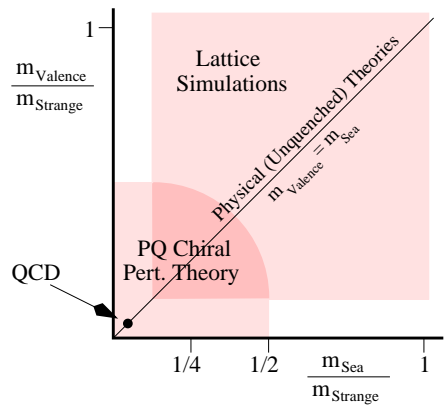}}
\caption{Parameter space of partially quenched theories.  Physical
theories, such as QCD, live on the diagonal line.  Lattice simulations
have been done with ``light" quark masses as low as $\sim
m_{strange}/8$.  Note that this plot is schematic, since there are
actually multiple valence and sea quarks, all of which can have
different masses.}\vspace{-0in} \label{fig:Plot}
\end{figure}

This situation can be improved using the partially quenched (PQ)
approximation, in which valence quarks (those which appear in external
states) and sea quarks (those which appear in loops) are allowed to
have different masses.  To ensure that valence quarks exist only in
external states, valence quarks have bosonic ghost quark partners of
equal mass. The presence of ghosts means that the PQ theory 
as a whole is unphysical, although the sea quark sector contained within 
it is physical.
Consideration of PQ theories expands the parameter space
available for lattice simulations and comparison to $\chi$PT, as shown
in Fig.~\ref{fig:Plot}.

In Refs.~\cite{PR,Phi0} it is argued that PQ QCD can be used 
for a quantitative
determination of the QCD low-energy constants.  
The primary assumption is that a generalization of $\chi$PT can
be used to describe the low-energy behavior of PQ QCD.
This theory, called PQ$\chi$PT, must contain all of
the operators of unquenched $\chi$PT in order to describe the behavior
of the unquenched sector.  Quark mass dependence in both $\chi$PT and
PQ$\chi$PT is explicit, so the coefficients in the Lagrangian depend
only on the number of quarks.  Therefore the
coefficients of operators in
the unquenched chiral Lagrangian are identical to those of
corresponding operators in the PQ chiral Lagrangian for $N_{sea} =
3$.\footnote{It is important for this argument that the $\eta'$ can be
integrated out of the PQ theory, which is shown in Ref.~\cite{Phi0}.}

To make use of this observation one fits results from PQ simulations
to the predictions of PQ$\chi$PT and thereby
determines the low-energy constants.  
Particular quantities studied in Refs.~\cite{SS,PR} were
next-to-leading order (NLO) PGB masses and decay constants, as well
as the hairpin vertex.  It was asserted for these calculations that
there are no new operators through NLO
in the PQ chiral Lagrangian, so that
no additional low-energy constants are needed at this
order.  Here we show that Refs.~\cite{PR,Phi0} missed one
operator in the NLO Lagrangian, and
explore the consequences.  While the existence of additional
operators does not fundamentally change the approach, it does
necessitate an elaboration of the program of using PQ QCD to determine
low energy constants.  

The general situation is as follows.
The PQ chiral Lagrangian contains two types of operators:
first, those which, in the sea quark sector, reduce to 
operators present in the QCD chiral Lagrangian, and whose coefficients
are thus identical to those in QCD,
and, second, operators which vanish in the sea quark sector.
We call the latter unphysical operators.
Their coefficients
are additional unphysical constants
needed to describe the low-energy behavior of the PQ theory.
They appear first at NLO in chiral perturbation theory,
with increasing numbers required as one works at higher order.
In general, both types of operator need to be taken into account
when matching
forms calculated in PQ$\chi$PT to PQ lattice data.  
It turns out, however, that for masses, decay constants and the
hairpin vertex, the new operators do not contribute until
next-to-next-to-leading order (NNLO).
Thus the results of Refs.~\cite{SS,PR} are unchanged.

This paper constitutes a preliminary investigation of the impact
of the unphysical operators on the utility of PQ QCD.
It is organized as follows.  In Sec.~\ref{sec:newop}, we
derive the existence and form of the new, linearly-independent
four-derivative operator present in the PQ chiral Lagrangian
at NLO.  We explore the
consequences of this operator 
in Sec.~\ref{sec:consqNLO} by discussing the type of meson
scattering processes to which it contributes.  
In Sec.~\ref{sec:consqNNLO} we continue the study
of Refs.~\cite{PR,Phi0} by calculating the leading contribution
of this operator to meson masses and decay constants,
which appear first at NNLO.  This raises the 
general issue of extending the approach of using PQ$\chi$PT to NNLO.
In Sec.~\ref{sec:analytic} we take another step in this direction by 
discussing the
analytic NNLO corrections to meson masses and decay constants.
We determine the operators that contribute, construct a minimal,
linearly-independent set, and show which of them vanish in the 
unquenched sector.
Using these operators, we determine the general form of NNLO analytic
meson mass and decay constant corrections in PQ$\chi$PT,
a result that is needed in fits to present lattice data.  
Finally, in Sec.~\ref{sec:conc}, we
summarize our findings, and discuss their general implications 
concerning the utility of PQ simulations.
       
\section{New Four-Derivative Operator in PQ$\chi$PT}
\label{sec:newop}

We first recall the chiral effective Lagrangian describing the 
properties of the light
pseudo-Goldstone boson (PGB) octet in QCD.  
Spontaneous breakdown of the approximate $SU(3)$ 
chiral symmetry of QCD by the vacuum,
\begin{equation}
	SU(3)_L \times SU(3)_R \rightarrow SU(3)_V
\,,
\end{equation}
leads to an octet of PGBs. 
They are conveniently collected into an $SU(3)$ matrix,
\begin{equation}
	\Sigma = e^{(2i\Phi/f)}
\end{equation}
\vspace{-.25in}
\begin{equation}
	\Phi = (\Phi_{ij}), \;\;\;\;\;\;\; i,j=u,d,s
\end{equation}
\vspace{-.25in}
\begin{equation}
	\Tr(\Phi)=0 \,,
\end{equation}
where $f$ is the leading-order meson decay constant.  
Under chiral symmetry transformations $\Sigma$ transforms as:
\begin{equation}
	\Sigma \rightarrow L \Sigma R^{\dagger} \label{eqn:trans}
\end{equation}
\vspace{-.25in}
\begin{equation}
 	L \in SU(3)_L, \;\;\;\;\;\; R \in SU(3)_R
\end{equation}
In the meson sector, $\chi$PT is an expansion in powers of 
$\epsilon^2 \sim m_{PGB}^2 / (4\pi f)^2 \sim p_{PGB}^2 / (4\pi f)^2$.  
The lowest-order Lagrangian is of $\cal{O}(\epsilon$$^2$$)$:
\begin{equation}	
{\cal{L}}_2 = \frac{{f}^2}{4}\Tr(\partial_\mu \Sigma \partial_\mu
\Sigma^{\dagger}) - \frac{{f}^2}{4}\Tr(\chi \Sigma^{\dagger} + \Sigma
\chi)
\label{eq:LOChL}
\end{equation}
Here $\chi$ is proportional to the quark mass matrix, 
\begin{equation}
	\chi_{ij}=\delta_{ij}\chi_i\,,
\end{equation}
and is defined such that the mass of 
a PGB containing quarks $i$ and $j$ is
\begin{equation}
	m_{ij}^2 = \frac{\chi_i + \chi_j}{2}
\label{eqn:LOMass}
\end{equation}
at lowest order.  Similarly, 
the $\cal{O}(\epsilon$$^4$$)$, or Gasser-Leutwyler, Lagrangian is\footnote{%
We do not consider 
the Wess-Zumino-Witten Lagrangian in this paper, 
since it does not contribute
to the processes we study until higher order than we consider.}
%
%
%
\begin{eqnarray}
	{\cal{L}}_4 & = & -L_1[\Tr(\partial_\mu \Sigma \partial_\mu
	\Sigma^{\dagger})]^2 - L_2 \Tr(\partial_\mu \Sigma
	\partial_\nu \Sigma^{\dagger}) \Tr(\partial_{\mu} \Sigma
	\partial_{\nu} \Sigma^{\dagger}) - L_3 \Tr(\partial_\mu \Sigma
	\partial_\mu \Sigma^{\dagger} \partial_\nu \Sigma \partial_\nu
	\Sigma^{\dagger}) \nonumber \\ & & + L_4 \Tr(\partial_\mu
	\Sigma \partial_\mu \Sigma^{\dagger})\Tr(\chi \Sigma^{\dagger}
	+ \Sigma \chi) + L_5 \Tr(\partial_\mu \Sigma \partial_\mu
	\Sigma^{\dagger}(\chi \Sigma^{\dagger} + \Sigma \chi)) - L_6
	[\Tr(\chi \Sigma^{\dagger} + \Sigma \chi)]^2 \nonumber \\ & &
	- L_7 [\Tr(\chi \Sigma^{\dagger} - \Sigma \chi)]^2 - L_8
	\Tr(\chi \Sigma^{\dagger} \chi \Sigma^{\dagger} + \Sigma \chi
	\Sigma \chi) + \mbox{contact terms}
\,.
\label{eqn:GL}
\end{eqnarray}
It is important to recall that $SU(3)$ group relations and
$\cal{O}(\epsilon$$^2$$)$ equations of motion were used to reduce the
operators in $\cal{L}$$_4$ to a minimal, linearly-independent basis.

\bigskip 
Partial quenching introduces unphysical, bosonic ghost quarks partners
of equal mass for the valence quarks \cite{BG}:
\begin{equation}
	\Sigma = e^{(2i\Phi/f)}
\end{equation}
\vspace{-.15in}
\begin{equation}
	\Phi = (\Phi_{ij}), \;\;\;\;\;\;\; i,j=
	\underbrace{A,B,\ldots}_{N_{valence}}, \:
	\underbrace{1,2,\ldots}_{N_{sea}}, \:
	\underbrace{\tilde{A},\tilde{B},\ldots}_{N_{valence}}
\end{equation}
\vspace{-.05in}
\begin{equation}
	\Str(\Phi)=0
\end{equation}
``Str" indicates supertrace, the graded analog of the trace.  The
transformation properties of $\Sigma$ are the same as in
Eq.~(\ref{eqn:trans}), except that the chiral symmetry group becomes a
graded group with both commutation and anti-commutation relations:
\begin{equation}
	SU(N_{sea}) \rightarrow SU(N_{valence} + N_{sea}|
	N_{valence}) \,.
\end{equation}
At leading order, the PQ chiral Lagrangian takes the same form
as for QCD, Eq.~(\ref{eq:LOChL}), except that traces are
replaced with supertraces. There are no additional operators
invariant under the new PQ chiral symmetry group.
It was asserted in Refs.~\cite{SS,PR,Phi0} that the same
is true for the NLO Lagrangian, $\cal{L}$$_4$.
This is in fact incorrect.  The number of four-derivative
operators in Eq.~(\ref{eqn:GL}) was reduced from four to three using
$SU(3)$ group relations.  This reduction does not occur in the PQ
theory, as we now explain.

\begin{figure}
	\epsfysize=0.8in
	\vspace{0in}\hspace{-.0in}\rotatebox{0}{\leavevmode\epsffile{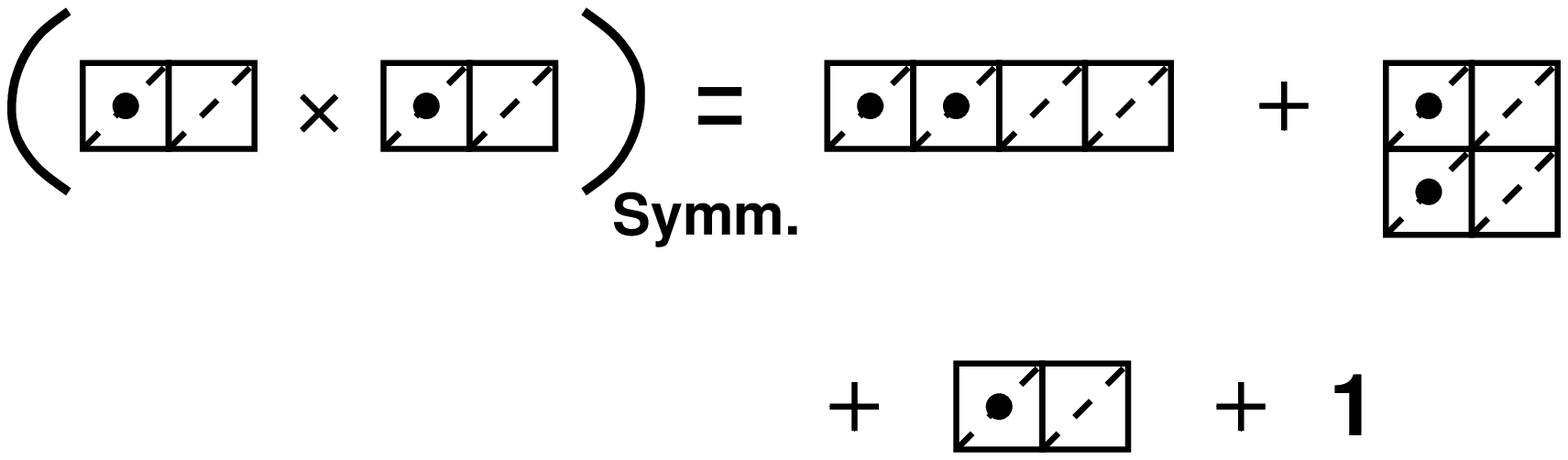}}
	\caption{Symmetric product of two adjoints in $SU(N | M)$.  The undotted
	boxes are fundamental representations and the dotted boxes are
	anti-fundamental representations.  The result also applies to $SU(N)$ for $N>3$ (with the dashed lines removed).}\vspace{-0in}
	\label{fig:GYT}
\end{figure}

We can use group theory to determine the number of
linearly independent four-derivative operators in 
${\cal L}_4$ in a general $SU(N |M)$ PQ theory.\footnote{%
In standard discussions for $SU(N)$, the analysis is usually
based on the Cayley-Hamilton theorem, from which one
can determine relations between products of traces of
matrices. This approach does not, however, generalize to graded groups,
so we use an alternative method which does generalize.}
It is useful to build such terms out the right-handed
Lie derivative\footnote{%
All terms involving derivatives in ${\cal L}_4$,
Eq.~(\ref{eqn:GL}),
can be written in terms of Lie derivatives
by appropriately inserting factors of $\Sigma^{\dagger}\Sigma = 1$.}
$\Sigma^{\dagger} \partial_{\mu} \Sigma$.
Under the chiral group this transforms as 
\begin{equation}
\left(\Sigma^{\dagger} \partial_{\mu} \Sigma\right) \longrightarrow
R \left(\Sigma^{\dagger} \partial_{\mu} \Sigma\right) R^{\dagger}
\,,
\end{equation}
and thus potentially contains an adjoint and a singlet component.
The singlet is absent because the Lie derivative has vanishing
supertrace. 
We can write it as
\begin{equation}
	\Sigma^{\dagger} \partial_{\mu} \Sigma = 
i \vec{l_{\mu}(\Phi(x))} \cdot 	\vec{\tau} 
\,,
\end{equation}
where $\tau$ is an $SU(N|M)$ generator,
and the vector $l_\mu$ is a real function of $\Phi$ and its
derivatives. The four-derivative terms contain four Lie derivatives,
with the Lorentz indices necessarily contracted in two pairs,
so that the terms have the schematic form
\begin{equation}
	(l_\mu\cdot\vec\tau) (l_\mu\cdot\vec\tau)
        (l_\nu\cdot\vec\tau) (l_\nu\cdot\vec\tau)
\,.
\end{equation}
Here group indices are not yet contracted,
and the issue is to determine the number of singlets 
under the chiral group that are
contained in this quantity.
The product 
$(l_\mu\cdot\vec\tau) (l_\mu\cdot\vec\tau)$ 
contains all representations arising in the symmetric
product of two adjoints.
Each of these can be combined to make a singlet
with the corresponding representation coming from
the other product with $\mu\to\nu$.
We conclude that the number of independent singlets 
that one can make out of right-handed Lie derivatives
is given by the number of representations contained 
in the symmetric product of two adjoints.
In fact, this gives all the independent 
four-derivative terms because all invariants that
can be built out of four left-handed, or
two left-handed and two right-handed, 
Lie derivatives can be rewritten
in terms of right-handed Lie derivatives alone, using
the cyclicity of the supertrace.

figure~\ref{fig:GYT} shows the symmetric product of two adjoints in
$SU(N | M)$.  We use the notation for graded Young tableaux of
Refs.~\cite{BB1,BB2}.  There are four representations, all of which
are self-conjugate.  They can each form a flavor singlet with the same
representation in the other pair of adjoints, so there are four
linearly-independent operators in the PQ theory.  Without the dashed
diagonal lines, Fig~\ref{fig:GYT} is also the symmetric product of two
adjoints in $SU(N)$ for $N > 3$.  However, 
in $SU(3)$, this product contains only
three representations:
\begin{equation}
	(8 \otimes 8)_{\mbox{Symm.}} = 27 \oplus 8 \oplus 1
\end{equation}
We conclude that there is one additional four-derivative operator in
the PQ chiral Lagrangian, just as there is an additional operator
in unquenched theories with four or more flavors.

It is straightforward to determine the form of the new operator.
In addition to the three in Eq.~(\ref{eqn:GL}) with trace replaced
by supertrace, we have:
\begin{equation}
	\mbox{Str}(\partial_\mu \Sigma \partial_\nu \Sigma^{\dagger}
	\partial^{\mu} \Sigma \partial^{\nu} \Sigma^{\dagger})
\end{equation}
For comparison with unquenched QCD, it is convenient to use a linear
combination of operators that vanish in the unquenched $SU(3)$ sector
of the PQ theory:\footnote{%
The fact that this combination vanishes in $SU(3)$ can be most easily
seen using the Cayley-Hamilton theorem, as nicely addressed in
Ref.~\cite{FS}.}$^,$\footnote{%
To be precise, $\cal{O}_{PQ}$ vanishes whenever $\Sigma$
has a block-diagonal form with one block, 
contained entirely within the ungraded $SU(N)$, being
an $SU(3)$ matrix, and the other being the identity.}
\vspace{-1pt}
\begin{eqnarray}
	\mathcal{O}_{PQ} & \! = \! & \Big\{ \Str(\partial_\mu
	\Sigma \partial_\nu \Sigma^{\dagger} \partial^{\mu} \Sigma
	\partial^{\nu} \Sigma^{\dagger}) -\frac{1}{2}
	\mbox{Str}(\partial_\mu \Sigma \partial^\mu
	\Sigma^{\dagger})^2 \nonumber \\ && - \mbox{Str}(\partial_\mu
	\Sigma \partial_\nu \Sigma^{\dagger})\cdot
	\mbox{Str}(\partial^{\mu} \Sigma \partial^{\nu}
	\Sigma^{\dagger}) + 2 \mbox{Str}(\partial_\mu \Sigma
	\partial^\mu \Sigma^{\dagger} \partial_\nu \Sigma \partial^\nu
	\Sigma^{\dagger}) \Big\}
\,.
\label{eqn:Op}
\end{eqnarray}
The operator $\cal{O}_{PQ}$ should be added to the
PQ chiral Lagrangian multiplied by
an undetermined coefficient, which we call $L_{PQ}$.
This new operator is the subtlety overlooked in
Refs.~\cite{PR,Phi0}.

\section{Consequences of the new operator at NLO}
\label{sec:consqNLO}

We now discuss the consequences of the additional operator.  In
general, its presence implies that PQ$\chi$PT has an additional NLO
coefficient that affects mesonic quantities, and which must be
accounted for in lattice fits.  Thus it complicates, but does not
invalidate, the program of using PQ$\chi$PT to extract the real
$\chi$PT coefficients.  In particular, since the operator
has four derivatives
it does not contribute to PGB masses and decay constants,
nor to the hairpin vertex of \cite{PR}, at tree-level.
It only contributes to these quantities at one-loop order,
which is NNLO in the chiral expansion. We return to these
contributions in the next section.

In this section we consider the quantities to which
$\cal{O}_{PQ}$ does contribute at NLO.
These are tree-level scattering processes, involving at least
four PGBs. We focus on the simplest, four-meson, scattering processes.
We illustrate such processes using quark line
diagrams, which show the flow of quarks within the
mesons.\footnote{Quark lines follow the flavor indices of the meson
fields within each supertrace.}  $\cal{O}_{PQ}$ contains both single
and double supertraces, so it contributes to both connected and
disconnected quark line diagrams.  Here we discuss the types of
charged meson scattering processes to which $\cal{O}_{PQ}$ does and
does not contribute.\footnote{We do not consider processes involving
neutral mesons because double poles in the meson propagators make it
unclear how to amputate diagrams and thus define scattering
amplitudes even at tree level.}

\begin{figure}
	\epsfysize=2.0in
	\vspace{0in}\hspace{-.0in}\rotatebox{0}
	{\leavevmode\epsffile{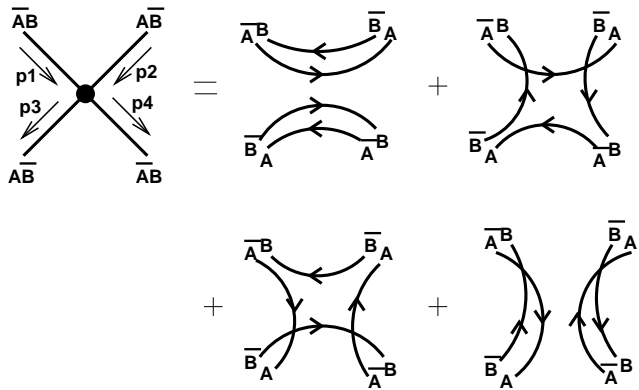}}
	\vspace{-0in}\caption{Scattering process to which
$\cal{O}_{PQ}$ does not contribute. 
It is physical when $A$ and $B$ are sea quarks,
and unphysical when they are valence quarks (with masses
differing from those of sea quarks).}
	\vspace{-0in}
	\label{fig:ABScatt}
\end{figure}

Because $\cal{O}_{PQ}$ vanishes in the unquenched $SU(3)$ limit, it
cannot contribute at tree-level to  scattering processes
that occur in the unquenched $SU(3)$ sector.  In fact,
since the operator has no dependence on the quark mass 
(i.e. no factors of $\chi$), it cannot distinguish between sea
and valence quarks. In other words, its contributions are the
same as in the chiral limit, in which limit there is a full
$SU(N_{valence}+N_{sea})$ symmetry among quarks.
It follows that {\em any} scattering process involving 
three or fewer quarks, whether they are valence or sea quarks,
is equivalent to an $SU(3)$ process, 
and cannot receive a contribution from $\cal{O}_{PQ}$. 
One such process is shown in
Fig.~\ref{fig:ABScatt}.  Mathematically, the contribution of
$\cal{O}_{PQ}$ to this process vanishes because relative minus signs
and numbers of Wick contractions among the quark line diagrams produce
cancellations.  Note that these cancellations would not occur if either
or both of the quarks in Fig.~\ref{fig:ABScatt} were bosonic.

\begin{figure}
	\epsfysize=.9in
	\vspace{-0in}\hspace{-.0in}\rotatebox{0}
	{\leavevmode\epsffile{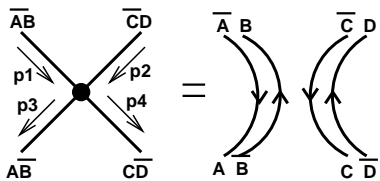}}
	\vspace{-0in}\caption{Unphysical scattering process
	because it involves more than three quarks. Here $A$, $B$, $C$, and $D$ are all different.}\vspace{-0in}
	\label{fig:4QuarkScatt}
\end{figure}

\begin{figure}
	\epsfysize=.95in
	\vspace{-0in}\hspace{-.0in}\rotatebox{0}
	{\leavevmode\epsffile{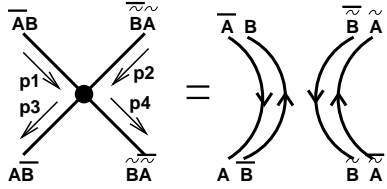}}
	\vspace{-0in}\caption{Unphysical scattering process
	because it involves ghost
	quarks.}\vspace{-0in} \label{fig:GhostScatt}
\end{figure}

$\cal{O}_{PQ}$ does contribute to scattering processes
that are unique to the PQ theory, and thus
unphysical.  For example, that shown in
Fig.~\ref{fig:4QuarkScatt} involves four quark flavors.  In the case
of PQ QCD with $N_{sea}=3$, these could be four valence quarks, three
sea quarks and a valence quark, or other possibilities.
figure~\ref{fig:GhostScatt} is also unique to the PQ theory, involving
both valence and ghost quarks.  It is clear that $\cal{O}_{PQ}$
contributes to these processes because there is only one quark-line
diagram and thus no possibility of cancellations.
These examples illustrate the origin of the additional operator:
the extra particles present in the PQ theory allow the
separation of individual quark contractions in a way not
possible in the unquenched sector.
The same explanation holds for physical theories with four or more
flavors.

These unphysical scattering processes can be used, in principle, to
determine $L_{PQ}$, the coefficient of the new operator in the PQ
chiral Lagrangian.  To illustrate this we have calculated 
some representative scattering amplitudes.
First, we give the expression for the ``physical''
process of Fig.~\ref{fig:ABScatt}:
\begin{eqnarray}
	\mathcal{M}_3 & \!\!\!\! = \!\!\!\! & -\frac{32}{f^4}L_1
	\Big[(p_1 \cdot p_3)(p_2 \cdot p_4) + (p_1 \cdot p_2)(p_3
	\cdot p_4)\Big] \nonumber \\ &&-\frac{16}{f^4}L_2 \Big[ (p_1
	\cdot p_2)(p_3 \cdot p_4) + (p_1 \cdot p_3)(p_2 \cdot p_4) +
	2(p_1 \cdot p_4)(p_2 \cdot p_3) \Big] \nonumber \\
	&&-\frac{16}{f^4}L_{3} \Big[((p_1 \cdot p_2)(p_3 \cdot p_4) +
	(p_1 \cdot p_3)(p_2 \cdot p_4) \Big] \nonumber \\
	&&-\frac{16}{f^4} L_4 m_{AB}^2 \Big[ (p_1 \cdot p_3) + (p_2 \cdot p_4) - (p_1
	\cdot p_2) - (p_3 \cdot p_4) \Big] \nonumber \\
	&& - \frac{4}{f^4}L_5 m_{AB}^2 \Big[(p_1 \cdot
	p_3) + (p_2 \cdot p_4) - (p_1 \cdot p_2) - (p_3 \cdot p_4)
	\Big] \nonumber \\ &&- \frac{64}{f^4} L_6 m_{AB}^4 -
	\frac{32}{3f^4} L_6 \left(\sum_{sea}\chi_{sea}\right)
	 m_{AB}^2 - \frac{128}{3f^4}
	L_8 m_{AB}^4 \label{eqn:M3}
\end{eqnarray}
where the momenta are Euclidean and are defined in the figure,
and $m_{AB}$ is the leading order mass as given in
Eq.~(\ref{eqn:LOMass}).  Note that, as claimed above,
the amplitude  does not contain a term
proportional to $L_{PQ}$. This is true even though the process
may involve only valence quarks with masses differing
from those of sea quarks, and thus be unphysical.

The unphysical processes in Figs.~\ref{fig:4QuarkScatt} 
and \ref{fig:GhostScatt} are chosen so that they only
receive contributions from disconnected quark line diagrams.
It follows that their
scattering amplitudes only receive contributions from double
supertrace operators, and both have terms proportional to $L_{PQ}$.
The amplitude for Fig.~\ref{fig:4QuarkScatt} is
\begin{eqnarray}
	\mathcal{M}_4 & \!\!\!\! = \!\!\!\! & -\frac{32}{f^4}L_1(p_1
	\cdot p_3)(p_2 \cdot p_4) -\frac{16}{f^4}L_2 \Big[ (p_1 \cdot
	p_2)(p_3 \cdot p_4) + (p_1 \cdot p_4)(p_2 \cdot p_3) \Big]
	\nonumber \\ &&-\frac{16}{f^4}L_{PQ} \Big[((p_1 \cdot p_2)(p_3
	\cdot p_4) + (p_1 \cdot p_3)(p_2 \cdot p_4) + (p_1 \cdot
	p_4)(p_2 \cdot p_3) \Big] \nonumber \\ &&-\frac{16}{f^4} L_4
	\Big[ m_{CD}^2(p_1 \cdot p_3) + m_{AB}^2(p_2 \cdot p_4) \Big]
	- \frac{32}{f^4} L_6 m_{AB}^2 m_{CD}^2, \nonumber \\
	&&-\frac{32}{f^4} L_6 \left(\sum_{sea}\chi_{sea}\right)
	\frac{(m_{AB}^2 + m_{CD}^2)}{2}\,,
	\label{eqn:M4}
\end{eqnarray}
The amplitude for  Fig.~\ref{fig:GhostScatt} is
really a special case of
Eq.~(\ref{eqn:M4}), obtained by setting $\chi_C=\chi_A$ and $\chi_D=\chi_B$,
and including an overall sign
because Fig.~\ref{fig:GhostScatt} contains one pair of bosonic quarks,
and therefore one supertrace over bosonic indices.  The kinematic
factors are identical because the quark line diagrams have the same
structure.  

These amplitudes provide a method, in principle, of determining
$L_{PQ}$ from PQ simulations. The idea is to calculate the
scattering amplitudes, as a function of the valence masses
and momenta, and thus determine $L_{1-3}$ from ${\cal M}_3$
($L_{4,5,6,8}$ having been obtained from fits to the PGB masses
at NLO~\cite{SS,PR}), and then determine $L_{PQ}$ from ${\cal M}_4$.
Other amplitudes involving different external quarks can also be
used. In practice, it is not possible to determine the scattering
amplitude itself, because the standard method for doing so relies
on unitarity~\cite{Luscher}, and this is violated at the
one-loop level in these amplitudes~\cite{BGscatt,Lin}.
Instead, what must be done is to use PQ$\chi$PT to predict the
form of Euclidean finite volume correlation functions,
which will be given in terms of $L_{PQ}$ and other low energy constants,
and fit this to lattice results for these correlation functions.\footnote{
The simplest example of this is the hairpin correlator discussed in Ref.~\cite{PR},
which is predicted to have a double pole, and thus is manifestly unphysical.
Nevertheless, its coefficient, if it can be determined, gives information about
physical low energy constants $L_5$ and $L_7$. We note that in the quenched theory,
where one also expects a double pole, its coefficient has been successfully 
determined from fits.}  
In this way the unphysical nature of the PQ theory is accounted for.

\section{Consequences of the new operator at NNLO}
\label{sec:consqNNLO}

\begin{figure}
	\epsfysize=.82in
	\vspace{0in}\hspace{-0in}\rotatebox{0}
	{\leavevmode\epsffile{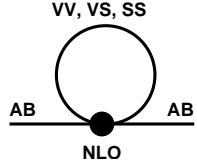}}
	\vspace{0in}\caption{PGB mass renormalization from
	$\cal{O}_{PQ}$ (represented by the solid circle). $V$ and $S$
	refer to valence and sea quarks respectively.}
	\vspace{0in} \label{fig:Mass}
\end{figure}

\begin{figure}
	\epsfysize=0.77in
	\vspace{-.1in}\hspace{0in}\rotatebox{0}
	{\leavevmode\epsffile{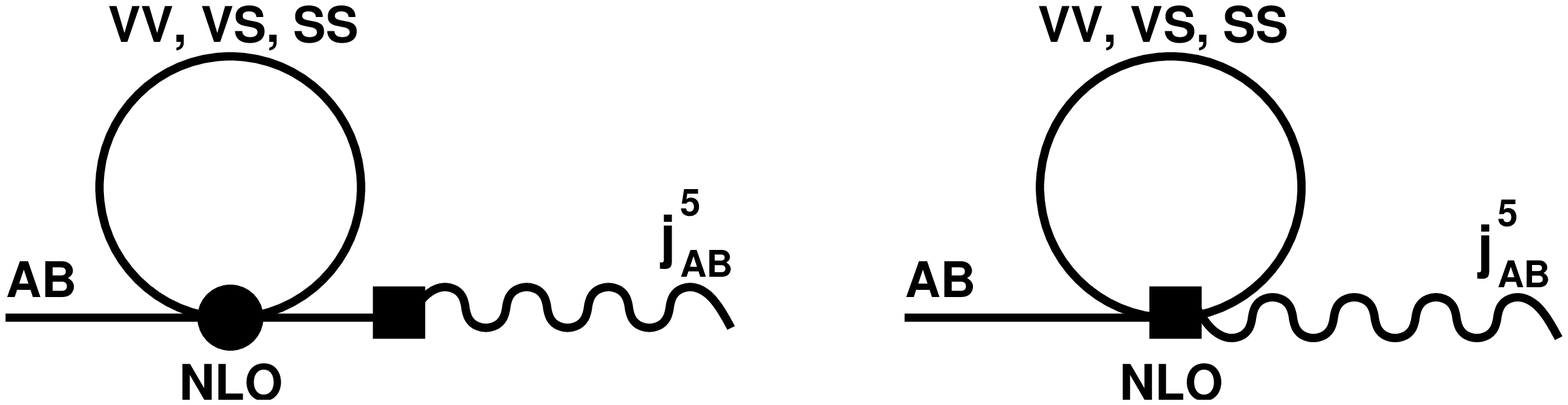}}
	\hspace{0in}\caption{PGB decay constant renormalization from
	$\cal{O}_{PQ}$. The filled square represents the axial current,
	while the filled circle represents interactions from the chiral Lagrangian.
	The label NLO indicates where $\cal{O}_{PQ}$ is contributing.
	}\vspace{0in}
\label{fig:Decay}
\end{figure}

\begin{figure}
	\epsfysize=2.6in
	\vspace{-.05in}\hspace{-.0in}\rotatebox{0}
	{\leavevmode\epsffile{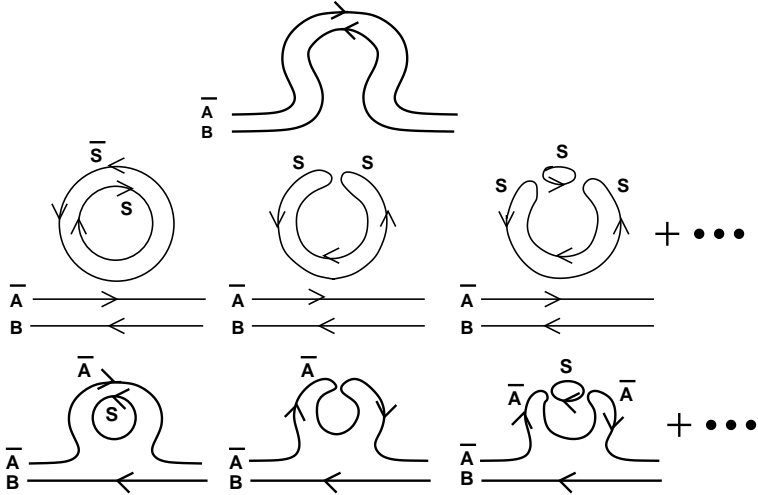}}
	\vspace{0in}\caption{Quark line diagrams contributing to mass and
	decay constant renormalization.}\vspace{0in} \label{fig:QL}
\end{figure}

As noted above, $\cal{O}_{PQ}$ contributes to
PGB masses and decay constants first at NNLO in PQ$\chi$PT.  
We have calculated these contributions as a first step in extending the 
application of PQ QCD to this order.
 To simplify calculations we
consider only two valence quarks, $A$ and $B$, and $N$ sea quarks of
equal mass.  We also only consider the properties of charged mesons,
$\pi_{AB}$.
figures~\ref{fig:Mass} and \ref{fig:Decay} show the
diagrams contributing to the mass and decay constant of charged mesons.
All of the meson diagrams receive
contributions from the classes of quark line diagrams shown in
figure~\ref{fig:QL}.  Quark line diagrams show the effect of partial
quenching---valence quarks appear only as external states because
ghost quarks cancel them in loops.  The series of sea quark hairpin
diagrams removes the flavor singlet propagator, enforcing the fact
that graded generators are traceless.

The corrections to the renormalized meson mass and decay constant from
the new PQ operator (using dimensional regularization and the
$\overline{MS}$ scheme of Ref.~\cite{DGH}) are:
\begin{eqnarray} 
	     { \Big( \delta {m}_{AB}^2 \Big) }_{L_{PQ}} & = & \Big(
	     \frac {\chi_A + \chi_B}{2} \Big)
	     \frac{L_{PQ}}{{16\pi}^2}\frac{8}{f^4} \times \Bigg\lbrace
	     (N^2 - 1){\chi_S}^2 \Big[\frac{-1}{8} -
	     \frac{3}{2}\ln{\chi_S}\Big] + \Big(\frac{\chi_A +
	     \chi_B}{2}\Big)^2 \Big[\frac{-1}{4}-3\ln{\frac{\chi_A +
	     \chi_B}{2\mu^2}}\Big] \nonumber \\ && + N
	     \Big(\frac{\chi_A + \chi_S}{2}\Big)^2
	     \Big[\frac{1}{4}+3\ln{\frac{\chi_A +
	     \chi_S}{2\mu^2}}\Big] + N \Big(\frac{\chi_B +
	     \chi_S}{2}\Big)^2 \Big[\frac{1}{4}+3\ln{\frac{\chi_B +
	     \chi_S}{2\mu^2}}\Big] \nonumber \\ && -
	     \frac{1}{N}\Big[{\chi_A}^2\Big(\frac{1}{4} + 3
	     \ln{\frac{\chi_A}{\mu^2}}\Big) + \chi_A\Big(\chi_S -
	     \chi_A\Big)\Big(-2 - 6 \ln{\frac{\chi_A}{\mu^2}}\Big)\Big]
	     \nonumber \\ && -
	     \frac{1}{N}\Big[{\chi_B}^2\Big(\frac{1}{4} + 3
	     \ln{\frac{\chi_B}{\mu^2}}\Big) + \chi_B\Big(\chi_S -
	     \chi_B\Big)\Big(-2 - 6 \ln{\frac{\chi_B}{\mu^2}}\Big)\Big]
	     \nonumber \\ && - \frac{2}{N}{\chi_A}^2 \Big(\frac{\chi_A
	     - \chi_S}{\chi_A - \chi_B}\Big) \Big[\frac{1}{8} +
	     \frac{3}{2} \ln{\frac{\chi_A}{\mu^2}}\Big] -
	     \frac{2}{N}{\chi_B}^2 \Big(\frac{\chi_B - \chi_S}{\chi_B
	     - \chi_A}\Big) \Big[\frac{1}{8} + \frac{3}{2}
	     \ln{\frac{\chi_B}{\mu^2}}\Big] \Bigg\rbrace
\label{eqn:Mass}
\end{eqnarray}

\begin{eqnarray} 
	     { \Big( \delta {f}_{AB}\Big) }_{L_{PQ}} & = & f
	     \frac{L_{PQ}}{{16\pi}^2}\frac{8}{f^4} \times \Bigg\lbrace
	     (N^2 - 1){\chi_S}^2 \Big[\frac{-1}{16} -
	     \frac{1}{4}\ln{\chi_S}\Big] + \Big(\frac{\chi_A +
	     \chi_B}{2}\Big)^2
	     \Big[\frac{-1}{8}-\frac{1}{2}\ln{\frac{\chi_A +
	     \chi_B}{2\mu^2}}\Big] \nonumber \\ && + N
	     \Big(\frac{\chi_A + \chi_S}{2}\Big)^2
	     \Big[\frac{1}{8}+\frac{1}{2}\ln{\frac{\chi_A +
	     \chi_S}{2\mu^2}}\Big] + N \Big(\frac{\chi_B +
	     \chi_S}{2}\Big)^2
	     \Big[\frac{1}{8}+\frac{1}{2}\ln{\frac{\chi_B +
	     \chi_S}{2\mu^2}}\Big] \nonumber \\ && +
	     \frac{1}{N}\Big[{\chi_A}^2\Big(\frac{-1}{8} -\frac{1}{2}
	     \ln{\frac{\chi_A}{\mu^2}}\Big) + \chi_A\Big(\chi_S -
	     \chi_A\Big)\Big(\ln{\frac{\chi_A}{\mu^2}}\Big)\Big]
	     \nonumber \\ && +
	     \frac{1}{N}\Big[{\chi_B}^2\Big(\frac{-1}{8} -\frac{1}{2}
	     \ln{\frac{\chi_B}{\mu^2}}\Big) + \chi_B\Big(\chi_S -
	     \chi_B\Big)\Big(\ln{\frac{\chi_B}{\mu^2}}\Big)\Big]
	     \nonumber \\ && + \frac{1}{N}{\chi_A}^2 \Big(\frac{\chi_A
	     - \chi_S}{\chi_A - \chi_B}\Big) \Big[\frac{-1}{8} -
	     \frac{1}{2} \ln{\frac{\chi_A}{\mu^2}}\Big] +
	     \frac{1}{N}{\chi_B}^2 \Big(\frac{\chi_B - \chi_S}{\chi_B
	     - \chi_A}\Big) \Big[\frac{-1}{8} - \frac{1}{2}
	     \ln{\frac{\chi_B}{\mu^2}}\Big] \Bigg\rbrace
\label{eqn:Decay}
\end{eqnarray}
where $\mu$ is the renormalization scale.  It can be seen that the
corrections in
Eqs.~\ref{eqn:Mass} and \ref{eqn:Decay} are NNLO by the fact that 
they have the form
$\chi^2\mbox{ln}(\chi)$.  A check of these
results is that if $A$ and $B$ are sea
quarks, and if $N_{sea} = 3$, the mesons in Figs.~\ref{fig:Mass} and
\ref{fig:Decay} live in the unquenched $SU(3)$ sector and cannot get
contributions from $\cal{O}_{PQ}$.  Equations~(\ref{eqn:Mass}) and
(\ref{eqn:Decay}) do, in fact, vanish in the unquenched limit: $\chi_A =
\chi_B = \chi_S$, $N_{sea}$ = 3.\footnote{The same holds true for
$N_{sea} = 2$, where $\cal{O}_{PQ}$ also vanishes.}

We stress that these results 
are only part of the NNLO charged
meson mass and decay constant corrections.  The full correction
receives 2-loop contributions from $\cal{L}$$_2$, 1-loop contributions
from terms in $\cal{L}$$_4$ other than $\cal{O}_{PQ}$, 
and tree-level contributions from $\cal{L}$$_6$.
We next discuss the contributions from $\cal{L}$$_6$.
    
\section{Analytic NNLO mass and decay constant corrections}
\label{sec:analytic}

Fits of present PQ lattice data require NNLO terms
\cite{Farchioni1,Farchioni2,Aubin}.  
Full non-analytic NNLO calculations in
PQ$\chi$PT are not available, so, as an intermediate step, we
have determined the form of the analytic NNLO mass and decay constant
corrections for charged mesons.
These arise from the tree-level diagram in Fig.~\ref{fig:Tree}:
\begin{figure}
	\epsfysize=0.32in
	\vspace{0in}\hspace{0in}
	\rotatebox{0}{\leavevmode\epsffile{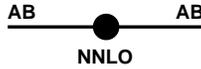}}
	\hspace{0in}\caption{NNLO PGB mass renormalization from
	operators in $\cal{L}$$_6$}\vspace{0in}
\label{fig:Tree}
\end{figure}

\noindent
This diagram gets contributions from operators in $\cal{L}$$_6$ with
zero and two derivatives.  Analytic, NNLO decay constant corrections
come from a similar diagram, but with an axial current insertion, and only come from two-derivative operators.  As in the case of
$\cal{O}_{PQ}$, we need to be sure to include all operators in the PQ
theory, some of which may be linearly-dependent in unquenched $SU(3)$.

First we determine the number of linearly-independent, derivative-free
operators in $\cal{L}$$_6$.  
As always, these must be invariant under chiral symmetry
transformations, hermitian, and invariant under the discrete
symmetries respected by QCD: C, P, and T.
They must have three $\chi$ matrices to
be $\cal{O}(\epsilon$$^6)$.  It is easy to construct terms with the
correct symmetry properties using the following two building
blocks with simple transformation properties:
\begin{eqnarray}
	\chi^\dagger \Sigma & \rightarrow &(R \chi^\dagger
	L^\dagger)(LUR^{\dagger}) = R(\chi^\dagger \Sigma)R^\dagger
	\nonumber \\ \Sigma^\dagger \chi & \rightarrow &(R
	\chi^\dagger L^\dagger)(L \chi R^{\dagger}) = R(\Sigma^\dagger
	\chi)R^\dagger \label{eqn:basis}
\end{eqnarray}
We find seven allowed forms:
\begin{enumerate}

\item{$\Big(\Str(\chi^\dagger \Sigma )\Big)^3 + \Big(\Str(\Sigma^\dagger \chi)\Big)^3$}

\item{$\Str(\chi^\dagger \Sigma \chi^\dagger \Sigma \chi^\dagger \Sigma ) + \Str(\Sigma^\dagger \chi \Sigma^\dagger \chi \Sigma^\dagger \chi)$}

\item{$\Str(\chi^\dagger \Sigma \chi^\dagger \Sigma )\Str(\chi^\dagger \Sigma ) + \Str(\Sigma^\dagger \chi \Sigma^\dagger \chi)\Str(\Sigma^\dagger \chi)$}

\item{$\Str(\chi^\dagger \Sigma \chi^\dagger \Sigma )\Str(\Sigma^\dagger \chi) + \Str(\Sigma^\dagger \chi \Sigma^\dagger \chi)\Str(\chi^\dagger \Sigma )$}

\item{$\Big(\Str(\chi^\dagger \Sigma )\Big)^2\Str(\Sigma^\dagger \chi) + \Big(\Str(\Sigma^\dagger \chi)\Big)^2\Str(\chi^\dagger \Sigma )$}

\item{$\Str(\chi^\dagger \Sigma \Sigma^\dagger \chi)\Str(\chi^\dagger \Sigma ) + \Str(\chi^\dagger \Sigma \Sigma^\dagger \chi)\Str(\Sigma^\dagger \chi)$}

\item{$\Str(\chi^\dagger \Sigma \Sigma^\dagger \chi \Sigma^\dagger \chi) + \Str(\Sigma^\dagger \chi \chi^\dagger \Sigma \chi^\dagger \Sigma ) $}

\end{enumerate}
(The last two can be simplified using $\Sigma^\dagger\Sigma=1$, but we choose
not to do so to better show their form.)
Our task is to reduce them to a linearly independent set
using group-specific relations.  

For $SU(3)$ we can use the Cayley-Hamilton theorem, which
states that any matrix satisfies its own characteristic equation, and
gives rise to numerous relations between traces of matrices.  The
relevant relation here
is that the determinant of a matrix can be expressed as a
polynomial in traces of the matrix of order the size of the matrix.
For $3 \times 3$ matrices, this polynomial is
\vspace{-.1in}\begin{equation}
	\det(M) = \frac{1}{3} \Tr(M^3) - \frac{1}{2} \Tr(M^2) \Tr(M) +
	\frac{1}{6}\Big(\Tr(M)\Big)^3 
\label{eqn:CH}
\end{equation}
Since $\det(\chi^\dagger\Sigma)=\det(\chi)^*=const.$,
the first three operators in the list
are linearly dependent.
Thus there are only six independent derivative-free $O(\epsilon^6)$ 
operators in $SU(3)$.

There is no 
analogous relationship among operators in the PQ theory, because
the super-determinant is not a finite polynomial~\cite{BB2}.  
The same holds true for
$SU(N)$, $N\ge 4$, since the Cayley-Hamilton relations involve
quartic or higher-order polynomials, 
and do not imply any linear dependence among third order polynomials.
We conclude there are seven derivative-free operators
in PQ$\chi$PT, and also in unquenched $SU(N)$ theories with
$N\ge 4$.  
One of the seven PQ operators can be chosen to be unphysical,
namely the linear combination given in Eq.~(\ref{eqn:CH}) with
trace replaced with supertrace.
While our PQ result is new, we note that
 Ref.~\cite{BCE} enumerates all of the operators in the
$\cal{O}(\epsilon$$^6$) chiral Lagrangian for unquenched $SU(N)$.
Ref.~\cite{BCE}
lists seven linearly-independent operators in $SU(N)$, $N\ge4$, and six
in $SU(3)$, so our results our consistent.

\begin{figure}
	\epsfysize=1.4in
	\vspace{0in}\hspace{-.0in}\rotatebox{0}{
	\leavevmode\epsffile{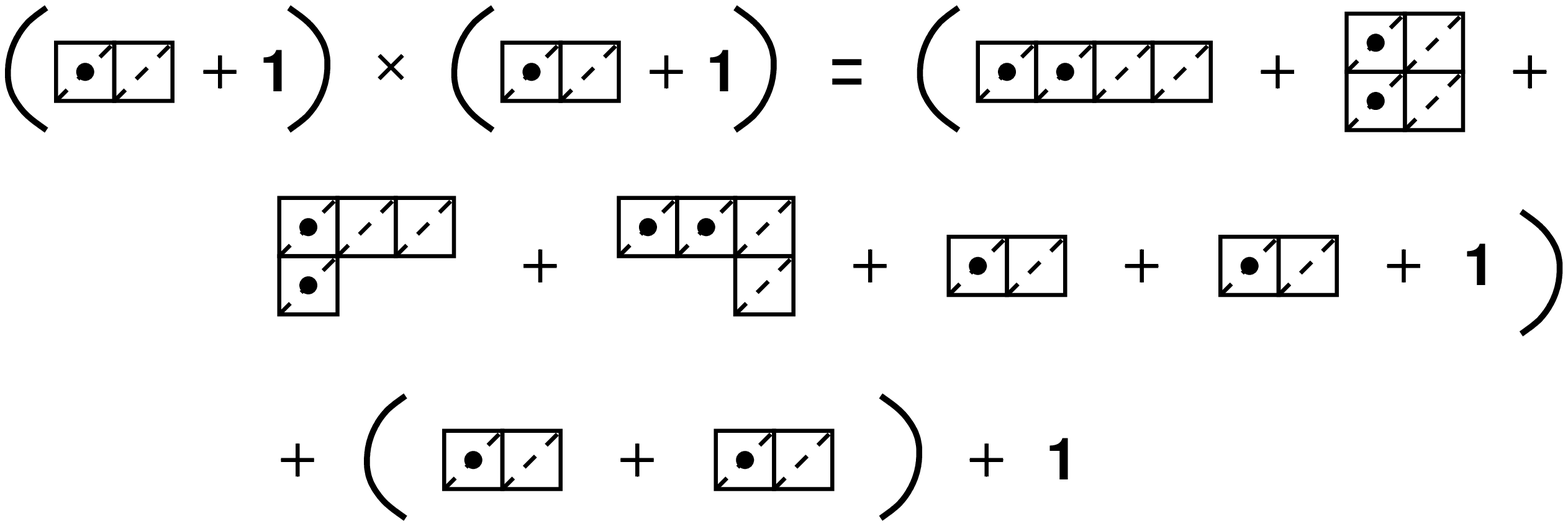}}
	\caption{The product of two bi-fundamentals in $SU(N | M)$.
	The result also applies for
	$SU(N)$ for $N > 3$ (with the dashed lines removed).}
	\vspace{-0in} \label{fig:bigGYT}
\end{figure}

\begin{figure}
	\epsfysize=1.05in
	\vspace{0in}\hspace{-.0in}\rotatebox{0}
	{\leavevmode\epsffile{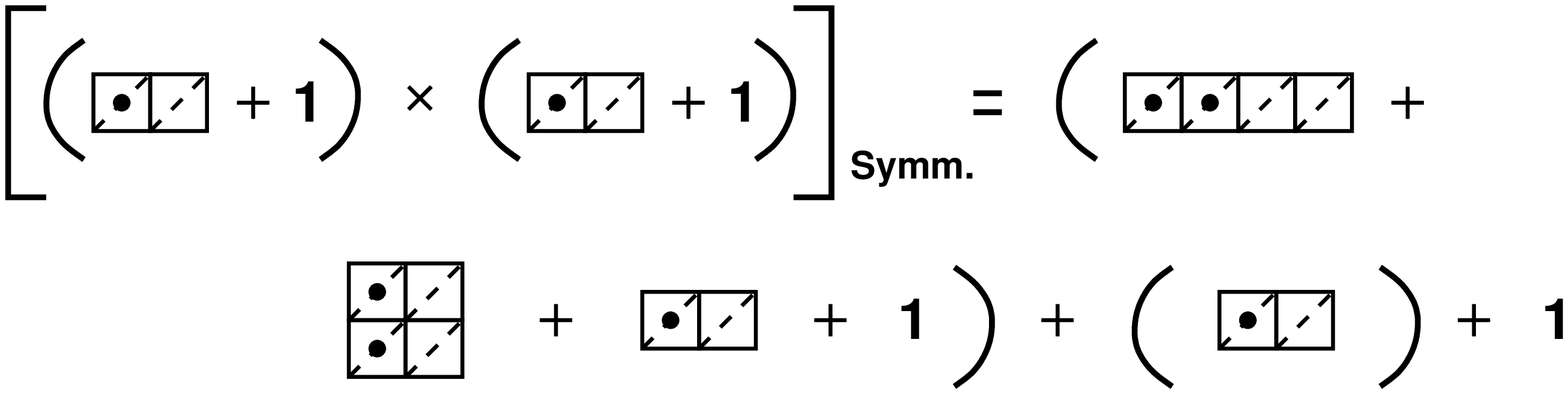}}
	\caption{Symmetric product of two bi-fundamentals
	in $SU(N | M)$.  The result also applies to
	$SU(N)$ for $N > 3$ (with the dashed lines removed).}
	\vspace{-0in} \label{fig:symGYT}
\end{figure}

For terms involving two derivatives, we can use group theory
to determine the number of independent operators, generalizing
the method used earlier in Sec.~\ref{sec:newop}.
To be $O(\epsilon^6)$ such operators must also contain two
factors of $\chi$.
We construct these operators using right-handed Lie derivatives and the matrices
defined in Eq.~(\ref{eqn:basis}), all of which transform
non-trivially only under the right-handed chiral group.    
We recall from Sec.~\ref{sec:newop}
that the Lie derivative transforms like an adjoint,
so an operator with two Lie derivatives (which are necessarily
contracted together by Lorentz symmetry)
must be symmetric in this pair of adjoints.
The symmetric product of two adjoints is shown in
Fig.~\ref{fig:GYT}. $\Sigma^\dagger \chi$ and $\chi^\dagger
\Sigma$ transform as bi-fundamentals under the
chiral symmetry group, but are not traceless, and so contain both
adjoint and singlet parts.  Therefore any operator that contains two
of these matrices also comes from the product of an adjoint plus
singlet times an adjoint plus singlet, as shown in
Fig.~\ref{fig:bigGYT}.  If the operator contains two of the same
matrices (either $\Sigma^\dagger \chi$ or $\chi^\dagger \Sigma$), it
will only come from the symmetric part of this product, shown in
Fig.~\ref{fig:symGYT}.  
Putting this together, we must 
construct singlets
out of the product in Fig.~\ref{fig:GYT}, which comes from the Lie
derivatives, and that in either Fig.~\ref{fig:bigGYT} or \ref{fig:symGYT},
which come from the terms involving factors of $\chi$.

In this way, we find that 
eight linearly-independent operators can be made out of two Lie
derivatives, one $\Sigma^\dagger \chi$, and one $\chi^\dagger \Sigma$.  
They correspond to the eight flavor singlets in the direct product of
the representations in Figs.~\ref{fig:GYT} and \ref{fig:bigGYT}.
One choice of basis is the following,
\begin{enumerate}

\item{$\Str(\partial_\mu \Sigma^\dagger \partial_\mu \Sigma \chi^\dagger \chi)$}

\item{$\Str(\partial_\mu \Sigma^\dagger \chi \chi^\dagger \partial_\mu \Sigma)$}

\item{$\Str(\partial_\mu \Sigma^\dagger \chi \Sigma^\dagger \partial_\mu \Sigma \chi^\dagger \Sigma)$}

\item{$\Str(\partial_\mu \Sigma^\dagger \partial_\mu \Sigma) \Str(\chi^\dagger \chi)$}

\item{$\Str(\partial_\mu \Sigma^\dagger \chi) \Str(\chi^\dagger \partial_\mu \Sigma)$}

\item{$\Str(\partial_\mu \Sigma^\dagger \partial_\mu \Sigma) \Str(\chi^\dagger \Sigma) \Str(\Sigma^\dagger \chi)$}

\item{$\Str(\partial_\mu \Sigma^\dagger \partial_\mu \Sigma \chi^\dagger \Sigma)\Str(\Sigma^\dagger \chi)$}

\item{$\Str(\partial_\mu \Sigma^\dagger \partial_\mu \Sigma \Sigma^\dagger \chi)\Str(\chi^\dagger \Sigma)$}

\end{enumerate}
where the derivatives act only on the quantity immediately to their right,
and we have used the anti-Hermiticity of the Lie derivatives to
simplify some of the operators.
Although group theory gives operators with the right chiral
transformation properties, we must impose the other symmetries by
hand.  Operators 1 and 2 transform into each other under parity, so we
must include their sum, with a single coefficient, in the Lagrangian.
Operators 7 and 8 are hermitian conjugates, so we must include their
sum as a single operator as well.  We conclude that
there are six operators
of the above type in the PQ chiral Lagrangian.

Similarly, six operators can be made out of two Lie derivatives and
two $\Sigma^\dagger \chi$s or $\chi^\dagger \Sigma$s because there are
two fewer adjoints in the symmetric product of Fig.~\ref{fig:symGYT}:
\begin{enumerate}

\item{$\Str(\partial_\mu \Sigma^\dagger \partial_\mu \Sigma \chi^\dagger \Sigma \chi^\dagger \Sigma + \partial_\mu \Sigma^\dagger \partial_\mu \Sigma \Sigma^\dagger \chi \Sigma^\dagger \chi))$}

\item{$\Str(\partial_\mu \Sigma^\dagger \chi \partial_\mu \Sigma^\dagger \chi + \chi^\dagger \partial_\mu \Sigma \chi^\dagger \partial_\mu \Sigma)$}

\item{$\Str(\partial_\mu \Sigma^\dagger \partial_\mu \Sigma) \Str(\chi^\dagger \Sigma \chi^\dagger \Sigma + \Sigma^\dagger \chi \Sigma^\dagger \chi)$}

\item{$\Str(\chi^\dagger \partial_\mu \Sigma)\Str(\chi^\dagger \partial_{\mu} \Sigma) + \Str(\partial_\mu \Sigma^\dagger \chi)\Str(\partial_\mu \Sigma^\dagger \chi)$}

\item{$\Str(\partial_\mu \Sigma^\dagger \partial_\mu \Sigma \chi^\dagger \Sigma) \Str(\chi^\dagger \Sigma) + \Str(\partial_\mu \Sigma^\dagger \partial_\mu \Sigma \Sigma^\dagger \chi)\Str(\Sigma^\dagger \chi)$}

\item{$\Str(\partial_\mu \Sigma^\dagger \partial_\mu \Sigma) \Str(\chi^\dagger \Sigma) \Str(\chi^\dagger \Sigma) + \Str(\partial_\mu \Sigma^\dagger \partial_\mu \Sigma) \Str(\Sigma^\dagger \chi) \Str(\Sigma^\dagger \chi)$}

\end{enumerate}

This completes the counting for the PQ theory: there are
twelve linearly independent two-derivative operators in
${\cal L}_6$. The same result applies to $SU(N)$ for $N>3$.
For $SU(3)$, however, one 
operator can be eliminated from each of the two previous lists in
because there is one fewer representation in the
product of adjoints for $SU(3)$ than for $SU(N|M)$.
We can use Cayley-Hamilton relations analogous to Eq.~(\ref{eqn:CH}) to 
obtain the
precise relationship between the operators.\footnote{See
Ref.~\cite{FS} for details.}  Thus there are only ten two-derivative
operators in $SU(3)$, and correspondingly two unphysical operators
in the PQ theory. Our results for $SU(N)$ are
consistent with those of Ref.~\cite{BCE}.

We stress that the counting of operators in all
$SU(N|M)$ PQ theories is valid for any $N>M\ge 2$.
The super-determinant is not a finite polynomial
in any such theory, and there are no Cayley-Hamilton
relations. From the perspective of Young tableaux, this result
follows because the theories do not have an antisymmetric
tensor to contract indices, and in particular there is
no limit to the number of 
boxes in the column of a Young tableau, as there is in $SU(N)$.

\bigskip
We now return to the NNLO corrections to the PGB mass and decay constant.
Although there are contributions from all of the operators 
in ${\cal L}_6$ enumerated above, the form
of the tree-level correction to the mass of $\pi_{AB}$ 
is simple:\footnote{%
This result has also been obtained independently by Ref.~\cite{Aubin}.}
\begin{eqnarray}
	{\Bigg( \!\frac{\delta m_{AB}^2}{m_{AB}^2}
	\!\Bigg)}_{\stackrel{NNLO,}{Analytic}} & = & \alpha_1 \tr(\chi_S^2)
	+ \alpha_2 \tr(\chi_S)^2 +
	+ \alpha_3 \tr(\chi_S) \Big( \chi_A + \chi_B \Big) + \alpha_4 \Big(
	\chi_A + \chi_B \Big)^2 + \alpha_5 \Big( \chi_A - \chi_B
	\Big)^2
	\,.
\label{eq:NNLOanalytic}
\end{eqnarray}
Here $\chi_S$ is the mass matrix of the sea quarks.
In fact, Eq.~(\ref{eq:NNLOanalytic}) 
is the most general quadratic polynomial separately symmetric in the
valence and sea quark masses.  The $\alpha_i$ are linear combinations of
$\mathcal{L}_6$ coefficients, and we have checked that they are independent,
so that no relations are predicted among the $\alpha_i$.
Corrections to $f_{AB}$ have the same
form with different coefficients.  Fits in Ref.~\cite{Farchioni2} to
PQ lattice data with $\frac{1}{3}m_S < m_{sea} < \frac{2}{3}m_S$ are
consistent with this NNLO formula.

It is interesting to consider the contributions to the $\alpha_i$
from the three unphysical operators. We find that
these operators do contribute to the $\alpha_i$ (in fact, to all five of them
in the basis we use), i.e. that the unphysical operators affect
PGB properties at tree level.\footnote{%
Of course, by definition,
the total contribution of unphysical operators to $\delta m_{AB}^2$ must
vanish when both $\chi_A$ and $\chi_B$ are equal to sea quark masses,
and we have checked this explicitly for the combination appearing
in Eq.~(\ref{eqn:CH}) with $N_{sea}=3$.}
This is different from ${\cal O}_{PQ}$, which does not affect PGB properties
until one-loop order. It implies that the $\alpha_i$ are linear combinations
of the coefficients of physical and unphysical operators.
Clearly, since the number of low energy constants entering at NNLO
(nineteen in the zero and two-derivative sector alone) exceeds the
number of constraints
from PGB masses and decay constants (ten in the PQ theory\footnote{There are five each from Eq.~(\ref{eq:NNLOanalytic}) and the analagous expression for $\delta f_{AB}$.}),
one must use other quantities in order to completely determine all the constants.

Just as at NLO~\cite{SS}, using the PQ theory can simplify this determination.
Consider the case of degenerate sea quarks.
The unquenched sector alone constrains only two combinations of
constants (one each from the single term proportional to $\chi_A^2=\chi_B^2=\chi_S^2$ in $\delta m_{AB}^2$ and $\delta f_{AB}$), although
these combinations necessarily involve only physical constants.
Consideration of the PQ theory adds three unphysical operators, but
this ``cost" is
 outweighed by the benefit of the six additional constraints
from Eq.~(\ref{eq:NNLOanalytic}).\footnote{There are three each from $\delta m_{AB}^2$ and $\delta f_{AB}$, rather than four, because
the first two terms collapse into a single term for degenerate sea quarks.}
In addition, using the PQ theory one
can add more values of $\chi_A$ and $\chi_B$ to the fit at relatively
small computational expense.

If one uses non-degenerate sea quarks then only the second benefit of
partial quenching applies. This is because, even in the unquenched sector, one can, in principle,
determine four of
the five $\alpha_i$ for both PGB masses and decay-constants,
using a number of choices for the sea quark masses.\footnote{%
With three sea quarks there are four independent
quadratic combinations of masses that are symmetric under 
$A\leftrightarrow B$: $(\chi_A\pm\chi_B)^2$, $\chi_C^2$ and 
$(\chi_A+\chi_B)\chi_C$.}
Moving to the PQ theory gives two more constraints but at the cost
of three unphysical operators.
Nevertheless, we suspect that this cost will be easily outweighed
by the additional data points that one can obtain in the PQ theory.

\section{Conclusion}
\label{sec:conc}

In this paper we have discussed a complication to the program of using
simulations of unphysical PQ theories to obtain physical low energy constants.
We find that there are operators in PQ$\chi$PT which vanish
when restricted to the physical, unquenched sector of the theory.
There is one such unphysical operator in the NLO Lagrangian ${\cal L}_4$,
and three in the zero and two derivative part of ${\cal L}_6$.
Generically, we expect the number of unphysical operators to increase
with the order of the calculation. Their presence is related to
the fact that correlation functions in the unquenched sector involve linear combinations
of quark contractions, while in the larger
PQ theory one can determine each contraction separately.

Dealing with this complication is, in principle, straightforward.
The contributions from unphysical operators can be disentangled from
those of physical operators as long as one considers enough quantities
in the fits. The extent of this complication will depend on the
example being considered. For the quantities we have studied in detail,
the PGB masses and decay constants, the new ${\cal O}(p^4)$
operator ${\cal O}_{PQ}$ does not contribute at tree level to PGB masses
and decay constants, and so one does not need to include it at all
until one works at NNLO.

As a case study in how to deal with the complication of unphysical
operators, we have outlined a strategy for how one could,
in principle, determine and subtract the NNLO contributions of ${\cal
O}_{PQ}$ to PGB properties. Our idea is to determine its coefficient
using unphysical scattering processes, to which ${\cal O}_{PQ}$
contributes at the first non-vanishing order, which is NLO in
$\chi$PT. Then, when doing a NNLO fit to the PGB properties, one can
subtract the contribution from ${\cal O}_{PQ}$, the form of which we
have calculated.  Of course, any such fit would require all other NNLO
terms, including two-loop contributions with vertices from ${\cal
L}_2$, and these are not yet available for the PQ theory.

We have also considered the more difficult question of how to separate the
contributions from the unphysical operators in ${\cal L}_6$.
The difference here is that these operators contribute to PGB masses and
decay constants at tree level, so their coefficients appear
in combination with those of the physical NNLO operators.
One must therefore rely on the general strategy of calculating enough
quantities to constrain all the coefficients, both physical and unphysical.

Clearly, extending the determination of low energy constants to NNLO 
in the meson sector is a significant undertaking, requiring extensive
$\chi$PT calculations beyond those presently available, and consideration
of many quantities in addition to PGB masses and decay constants.
If such a program is undertaken, PQ simulations can play an important role.
While they introduce a few additional unknown constants, they
provide both additional constraints on low energy constants from the extra functions
of valence masses that are available, and, most importantly, the possibility
of additional, relatively cheap, data points to include in the fits.

Finally, we take this opportunity
to reiterate the purpose of simulating PQ QCD.
The intent is \emph{not} to study it as an approximation to,
or a model of, QCD, 
for it is an unphysical theory, but rather to use it as a tool to more
easily extract the low energy constants of QCD.
Once these are determined, they can be used to
calculate PGB processes in QCD, including those that are not
easily accessible to lattice calculations.
The unphysical nature of PQ QCD is  manifested in several 
ways, for example by the presence of double poles in neutral propagators, 
and by the inability to
define decay and scattering amplitudes~\cite{Lin}.
We have discussed here another unphysical feature, namely the presence
of additional operators in PQ$\chi$PT which do not contribute in
the unquenched sector.\footnote{%
A similar phenomena occurs in the calculation of weak matrix
elements~\cite{GoltermanPallante}.}
While all these features do complicate the use of PQ QCD,
we have argued that these complications can be overcome
and that PQ simulations remain an important element of the
lattice practitioner's toolkit.

\section*{Acknowledgments}
We thank Maarten Golterman,
David Lin, Guido Martinelli and Chris Sachrajda for useful
discussions.
This work was supported in part by the US Department of Energy through
grant DE-FG03-96ER40956/A006.

\end{document}